\documentclass{appolb}
\usepackage{graphicx}
\usepackage{amsfonts,amsmath}
\usepackage{bigstrut}
\newcommand{\hc}{\text{h.c.}}
\newcommand{\lt}{\mathcal{L}} 
\newcommand{\sul}[1]{\left( #1 \right)}

\newcommand{\inv}[1]{\frac{1}{#1}}
\newcommand{\eps}{\varepsilon}
\def \be#1\ee{\begin{equation}#1\end{equation}}
\def\bsp#1\esp{\begin{split}#1\end{split}}
\def\bal#1\eal{\begin{align}#1\end{align}}
\newcommand{\ri}{\text{i}} 
\newcommand{\tri}[3]{\begin{pmatrix}
        #1 \\ #2 \\ #3
\end{pmatrix}}

\begin{document}
    \title{Super-weakly coupled U(1)$_z$ and GeV neutrinos%
        \thanks{Presented at Matter to the Deepest 2021 conference.}%
    }
    \author{Timo J. Kärkkäinen$^1$ and Zoltán Trócsányi$^{1,2}$
        \vspace{0.5cm}        
        \address{$^1$Institute for Theoretical Physics, ELTE Eötvös Loránd University, Pázmány Péter sétány 1/A, 1117 Budapest, Hungary}
        \address{$^2$ ELKH-DE Particle Physics Research Group,
            4010 Debrecen, PO Box 105, Hungary}
        }

    \maketitle
    \begin{abstract}
        The \textit{super-weak force} combines three simple extensions of the Standard Model, one in gauge sector, one in fermion sector and one in scalar sector. Each of these extensions are well motivated by their rich phenomenology. Combined to a single framework, they can explain several open questions in particle physics and cosmology: the origin of dark matter, cosmic inflation, matter-antimatter asymmetry, neutrino masses and vacuum stability. We discuss the effects of the model on neutrino masses and phenomenology in the case where the heaviest sterile neutrinos have a GeV scale mass.
    \end{abstract}
    
    \section{Introduction}
    The incompleteness of the standard model (SM) of particle physics, when confronted with Nature, is a well-known fact. While the SM performs suspiciously well on the description of particle interactions, the discovery of neutrino oscillations in vacuum and matter - among other phenomena - has made abundantly clear that the SM must be extended to include new interactions. Even simple extensions may lead to a rich phenomenology. These include extensions on the scalar, fermion and gauge sectors. One could extend the model by an extra singlet scalar boson, a heavy neutral lepton or by a small gauge group. All three of these possibilities are taken into account and combined in a single framework which is called the \textit{super-weak model} \cite{Trocsanyi:2018bkm}.
    
    \section{Super-weak model}
    
\begin{figure}[htb]
        \centerline{%
            \includegraphics[width=12.5cm]{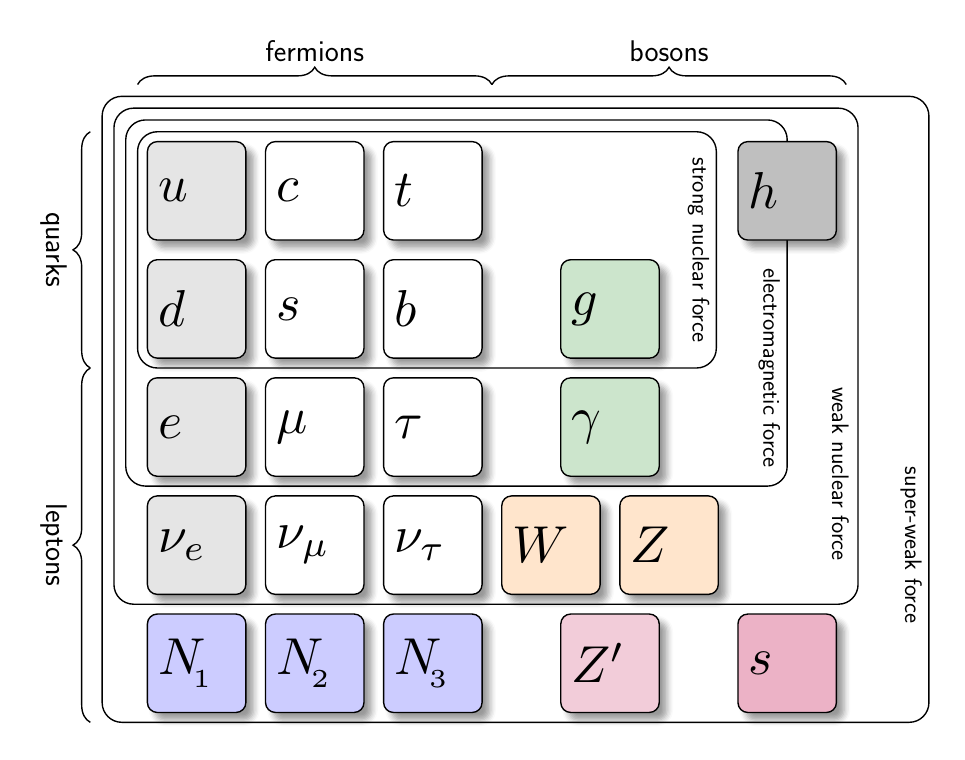}
}
        \caption{Particle content of super-weak model. The forces act on all particles within their respective boxes.}
        \label{Fig:Particles}
    \end{figure}

    The super-weak model includes three simple extensions of the SM: one on gauge sector, one on scalar sector and one on fermion sector. The fields are presented in a diagram in Fig. \ref{Fig:Particles}, and their gauge group representations and charges in Table \ref{tab:charges}.
    
    \begin{table} 
        \centering 
        \begin{tabular}{|c|ccc|}\hline 
            \textbf{Field} &     SU(2)$_L$ & U(1)$_Y$ & U(1)$_Z$ \\ \hline 
            \rule{0pt}{3ex}$Q_L$      &  \textbf{2} & $\frac{1}{6}$& $\frac{1}{6}$\\\hline 
            \rule{0pt}{3ex}$u_R$       &  \textbf{1} & $\frac{2}{3}$& $\frac{7}{6}$\\\hline 
            \rule{0pt}{3ex}$d_R$       &  \textbf{1} & $-\frac{1}{3}$& $-\frac{5}{6}$\\\hline 
            \rule{0pt}{3ex}$L_L$       &  \textbf{2} & $-\half$& $-\frac{1}{2}$\\\hline 
            \rule{0pt}{3ex}$\ell_R$    & \textbf{1} & $-1$& $-\frac{3}{2}$\\\hline 
            \rule{0pt}{3ex}$\phi$      & \textbf{2} & $\half$& 1\\\hline 
            \rule{0pt}{3ex}$N_R$       & \textbf{1} & $0$& $\frac{1}{2}$\\\hline 
            \rule{0pt}{3ex}$\chi$     & \textbf{1} & $0$&  $-1$\\\hline 
        \end{tabular}
        \caption{\label{tab:charges} Gauge group representations and charges of the fermions and scalar bosons of the super-weak model.}
    \end{table}
    \subsection{Gauge extension}
    
    The gauge group of the super-weak model is
    \begin{equation}
        \text{SU}(3)_\text{c} \otimes \text{SU}(2)_\text{L} \otimes \text{U}(1)_Y \otimes \text{U}(1)_z,
    \end{equation}
    that is, the SM group is extended by an extra U(1) group. The kinetic terms of the U(1)$_Y\otimes$U(1)$_z$ sector of the group can be described with the Lagrangian density
    \be
    \lt^\text{U(1)} = -\inv{4}F^{\mu\nu}F_{\mu\nu} - \inv{4}F^{\prime\mu\nu}F'_{\mu\nu} - \frac{\eps}{2}F^{\mu\nu}F'_{\mu\nu},
    \ee 
    where $F_{\mu\nu}$ and $F'_{\mu\nu}$ correspond to the field strength tensors of U(1)$_Y$ and U(1)$_z$. The model exhibits kinetic mixing, driven by a small real dimensionless coupling $\eps$. The covariant derivative acting on the fermion field $f$ can be written as
    \be
    \mathcal{D}_\mu^{U(1)} = \partial_\mu -\ri (y^fg_yB_\mu + z^fg_zB'_\mu),
    \ee 
    when we suppress the non-Abelian contribution. The $y^f$ and $z^f$ are the hypercharge and U(1)$_z$ charge of $f$. Equivalently we may choose a basis where the kinetic mixing is absent. Then, the covariant derivative can be written as
    \be 
    \mathcal{D}_\mu^{U(1)} = \partial_\mu - \ri (y,z)\begin{pmatrix}
        g_{y} & -\eta g_z' \\ 0 & g_z'
    \end{pmatrix}
    \begin{pmatrix}
        \cos \eps' & \sin \eps' \\ -\sin \eps' & \cos \eps'
    \end{pmatrix}\binom{\hat{B}_\mu}{\hat{B'}_\mu},
    \ee 
    where we have denoted $\eta = \eps g_y/g_z$ and $g_z' = g_z/\sqrt{1-\eps^2}$. The rotation angle $\eps'$ is not physical, since it can be absorbed to re-defined gauge fields. The mass eigenstates ($A_\mu,Z_\mu,Z'_\mu)$ are related to these fields via a rotation,
    \be
    \tri{\hat{B}_\mu}{W^3_\mu}{\hat{B'}_\mu} = \begin{pmatrix}
        \cos \theta_W & -\cos \theta_Z \sin \theta_W & -\sin \theta_Z \sin \theta_W \\
        \sin \theta_W & \cos \theta_Z \cos \theta_W & \cos \theta_W \sin \theta_Z \\
        0 & -\sin \theta_Z & \cos \theta_Z
    \end{pmatrix}\tri{A_\mu}{Z_\mu}{Z'_\mu},
    \ee 
    where $\theta_W$ is the weak mixing angle and $\theta_Z$ is the $Z-Z'$ mixing angle. The interactions of a neutral vector boson $V=Z$ or $Z'$ with fermion $f$ can be written as a chiral decomposition:
    \be
    \Gamma^\mu_{Vff} = -\ri e\gamma^\mu \sul{C^R_{Vff}P_R + C^L_{Vff}P_L},\quad P_{L,R} = \half (\textbf{1} \mp \gamma_5).
    \ee 
    The coefficients corresponding to the coupling of a vector boson to chiral fermion fields can be written in a simple form of rotation:
    
    \begin{align}\label{eq:eC}
        -e\binom{C^{L,R}_{Zff}}{C^{L,R}_{Z'ff}} &= 
        \begin{pmatrix}
            \cos \theta_Z & \sin \theta_Z \\ -\sin \theta_Z & \cos \theta_Z
        \end{pmatrix}
        \binom{T_3^f + Q^f\sin^2\theta_W}{-(y^fg_y' + z^fg_z')}
    \end{align}
    Clearly the coefficient corresponding to $Z'$ interaction can be obtained from $Z$ interaction by simply transforming $(\sin \theta_Z, \cos \theta_Z) \mapsto (\cos \theta_Z, -\sin \theta_Z)$. The factors $T_3^f$ and $Q^f$ are the eigenvalues of third SU(2) operator and electric charge in units of elementary charge.
    
    \begin{table} 
        \centering 
        \begin{tabular}{|c||c|c|c|c|}\hline \hline 
            $f$ &$y$ & $z$ & $T_3$ & $Q$ \\\hline \hline 
            \rule{0pt}{3ex}$u_L$   & $-\frac{1}{6}$ & $\frac{1}{6}$ & $-\half$ & $\frac{2}{3}$ \bigstrut \\\hline 
            \rule{0pt}{3ex}$u_R$   & $-\frac{2}{3}$ & $\frac{7}{6}$ & 0 & $\frac{2}{3}$ \bigstrut\\\hline 
            \rule{0pt}{3ex}$d_L$   & $-\frac{1}{6}$ & $\frac{1}{6}$ & $\half$ & $-\frac{1}{3}$ \bigstrut\\\hline  
            \rule{0pt}{3ex}$d_R$   & $\frac{1}{3}$ & $-\frac{5}{6}$ & 0 & $-\frac{1}{3}$ \bigstrut\\\hline 
            \rule{0pt}{3ex}$\nu_L$ & $\half$ & $-\half$ & $-\half$ & 0 \bigstrut\\\hline 
            \rule{0pt}{3ex}$N_R$   & 0 & $\half$ & 0 & 0 \bigstrut\\\hline 
            \rule{0pt}{3ex}$\ell_L$& $\half$ & $-\half$ & $\half$ & $-1$ \bigstrut\\\hline 
            \rule{0pt}{3ex}$\ell_R$& 1 & $-\frac{3}{2}$ & 0 & $-1$\bigstrut\\\hline \hline 
        \end{tabular}
        \caption{\label{tab:couplings} Eigenvalues of the U(1) charge operators, third SU(2)$_L$  generator and electric charge operator corresponding to chiral fermions of the super-weak model.}
    \end{table}
    
    \subsection{Scalar extension}
    
    The scalar sector of the super-weak model consists of the SM Higgs SU(2) doublet $\phi$ with charges $(y_\phi, z_\phi) = (1/2,1)$ and a complex singlet scalar $\chi$ with charges $(y_\chi, z_\chi) = (0,-1)$. The relevant Lagrangian corresponding to them is
    \be 
    \lt_\text{scalar} = |D_\mu\phi|^2 + |D_\mu\chi|^2 - \mu_\phi^2|\phi|^2 - \mu_\chi^2|\chi|^2 - \lambda_\phi|\phi|^4 - \lambda_\chi|\chi|^4 - \lambda|\phi|^2|\chi|^2,
    \ee 
    and we parametrize the fields after spontaneous symmetry breaking (SSB) in $R_\xi$ gauge as
    \be 
    \phi = \inv{\sqrt{2}}\binom{-i\sqrt{2}\sigma^+}{v+h'+i\sigma_\phi},\quad \chi = \inv{\sqrt{2}}\sul{w+ s'+i\sigma_\chi},
    \ee 
    where $v \simeq  246.22$ GeV and $w$ are the vacuum expectation values  and the fields $h',s',\sigma_\chi$ and $\sigma_\phi$ are real. The fields $\sigma_\phi$ and $\sigma_\chi$ correspond to the Goldstone bosons. To obtain physical fields, we perform standard field rotations
    \begin{align} 
        \binom{h}{s} = Z_S(\theta_S)\binom{h'}{s'},\quad \binom{\sigma_{Z}}{\sigma_{Z'}} = Z_G(\theta_G)\binom{\sigma_\phi}{\sigma_\chi },\quad Z_X = \begin{pmatrix}
            \cos \theta_X & \sin \theta_X \\ -\sin \theta_X & \cos \theta_X
        \end{pmatrix}
    \end{align} 
    where $\theta_S$ and $\theta_G$ are the scalar and Goldstone mixing angles, given by the relations
    \bal
    \sin \theta_S &= -\frac{\lambda vw}{\lambda_\phi v^2 - \lambda_\chi w^2},& \tan \theta_G &= \frac{M_{Z'}}{M_Z}\tan\theta_Z.
    \eal 
    
    \subsection{Fermion extension}
    
    The fermion sector of the super-weak model is extended with three sterile massive Majorana neutrinos $N_R = (\nu_4,\nu_5,\nu_6)$. We denote the active SM neutrinos as $\nu_L = (\nu_e,\nu_\mu,\nu_\tau)$ in the flavor basis and $(\nu_1,\nu_2,\nu_3)$ in the mass basis. The Majorana mass term corresponding to $N_R$ can not be directly included in the Lagrangian, since $N_R$ has a nonzero U(1)$_z$ charge. Instead, such a mass term is generated dynamically by SSB. The gauge invariant Yukawa interactions of the neutrinos are given by Lagrangian density
    \be
    \lt_Y^\nu = -\overline{N_R}Y_\nu \eps_{\alpha\beta} L_{L\alpha}\phi_\beta -\half\overline{N_R}Y_N(N_R)^c\chi + \hc,
    \ee 
    where $\alpha$ and $\beta$ are SU(2)$_L$ indices and $\eps_{\alpha\beta} = \begin{pmatrix}
        0 & 1 \\ -1 & 0
    \end{pmatrix}$ . After SSB, the neutrino mass terms and neutrino-scalar interaction terms are generated. Defining the $3\times 3$ Dirac and Majorana mass matrices
    \be 
    M_D = \frac{v}{\sqrt{2}}Y_\nu,\quad M_N = \frac{w}{\sqrt{2}}Y_N,
    \ee 
    the neutrino mass terms can be collected in a form
    \be 
    \lt_m^\nu = -\half\sul{\nu_L,\:(N_R)^c}C\begin{pmatrix}
        0 & M_D^T \\ M_D & M_N
    \end{pmatrix}\binom{\nu_L}{(N_R)^c},
    \ee 
    which has the exact form of type-I seesaw mechanism. The light neutrino mass matrix $M_L = -M_DM_N^{-1}M_D^\dagger + \hc $ can be obtained by block-diagnonalizing the full $6 \times 6$ neutrino mass matrix $M$ via a unitary matrix $U$:
    \be 
    U^TMU = U^T\begin{pmatrix}
        0 & M_D^T \\ M_D & M_N
    \end{pmatrix}U = M_\text{diag}=\text{diag}(m_1,\dots,m_6),
    \ee 
    where the masses $m_1,\dots,m_6$ correspond to neutrinos $\nu_1,\dots,\nu_6$. Writing the diagonalizing matrix as
    \be 
    U = \binom{U_L}{U_R^*},
    \ee 
    where $U_L$ and $U_R$ are semiunitary $3\times 6$ matrices, it is straightforward to derive the conditions
    \be 
    U_LU_L^\dagger = U_RU_R^\dagger = \textbf{1}_3,\quad U^\dagger U = U_L^\dagger U_L + U_R^TU_R^* = \textbf{1}_6
    \ee 
    for them.
    
    \section{Neutrino physics in super-weak model}
    
    \subsection{Radiative corrections to neutrino masses}
    
    \begin{figure}[th]
        \begin{center}
            %
            
            \begin{tabular}{|c|c|}\hline 
                \rule{0pt}{18ex}\includegraphics[width=0.47\linewidth]{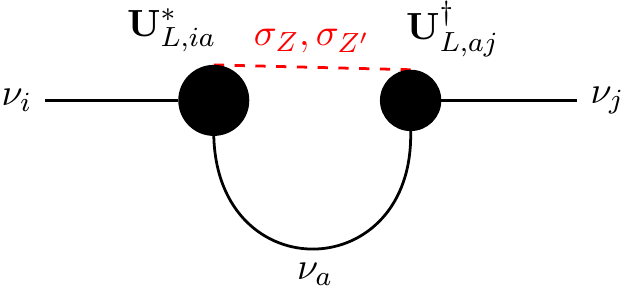}&
                \includegraphics[width=0.47\linewidth]{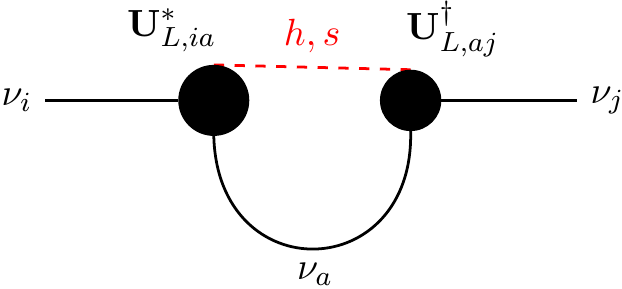}\\\hline 
                
                \rule{0pt}{18ex}\includegraphics[width=0.47\linewidth]{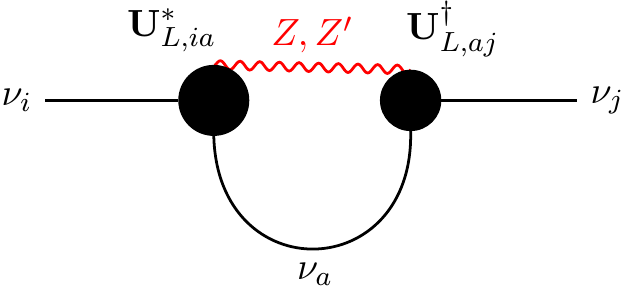}&
                \includegraphics[width=0.47\linewidth]{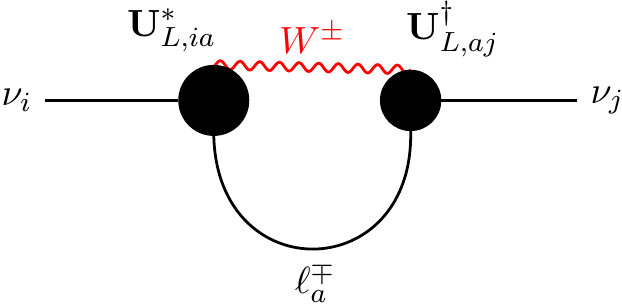}\\\hline 
            \end{tabular}
        \end{center}
        \caption{\label{fig:loop}
            Relevant one-loop Feynman diagrams contributing to the neutrino self energy. Clockwise from top left: Goldstone boson, scalar boson, charged gauge boson and neutral gauge boson contributions in super-weak model. The diagram containing $W$ boson does not contribute to the 
            matrix $\delta M_L$.}
    \end{figure}
    
    The loop-induced corrections to light neutrino masses receive additional contributions in the super-weak model compared to the standard seesaw scenario. \textit{A priori} it is not known whether the mass scales which are phenomenologically viable are consistent with small corrections to $M_L$. In Ref. \cite{Iwamoto:2021wko}, we have calculated the one-loop corrections to light neutrino mass matrix:
    \be 
    \delta M_L = \inv{16\pi^2}\sum \limits_{k=1,2}\sul{ 3(Z_G)_{k1}\frac{M_{V_k}}{v^2}F\sul{M_{V_k}^2} + (Z_S)_{k1}\frac{M_{S_k}}{v^2}F\sul{M_{S_k}^2}}
    \ee 
    Here we have denoted $(V_1,V_2) = (Z,Z')$ and $(S_1,S_2) = (h,s)$. The $3 \times 3$ matrix valued function $F$ is defined as
    \be 
    F_{ij} \equiv F_{ij}\sul{M^2;U_L,\lbrace m_a \rbrace_{a=1}^6} = \sum \limits_{a=1}^6 (U_L^*)_{ia}(U_L^\dagger)_{aj}m_a\sul{\frac{m_a}{M}}^2\frac{\ln \frac{m_a^2}{M^2}}{\frac{m_a^2}{M^2}-1}
    \ee 
    In the super-weak model, the mass scales of the new particles $\nu_4, \nu_5, \nu_6, s$ and $Z'$ are chosen as in Table \ref{tab:masses}. We estimate the elements of $\delta M_L$ on these mass scales, and then the eigenvalues of the full neutrino mass matrix $M_L + \delta M_L$ can be calculated. We find that the corrections to tree-level masses in the super-weak region of the parameter space are at most $\mathcal{O}(0.1)$~\%.
    
    \begin{table}[]
        \centering 
        \begin{tabular}{|c|c|c|}
            \hline
            \textbf{Particle} & \textbf{Freeze-in DM} & \textbf{Freeze-out DM} \\ \hline
            $\nu_4$ & $\mathcal{O}(10)$ keV & $\mathcal{O}(10)$ MeV  \\ \hline
            $\nu_5,\nu_6$ & \multicolumn{2}{c|}{$\mathcal{O}(1)$ GeV}    \\ \hline
            $s$ & \multicolumn{2}{c|}{$\mathcal{O}(100)$ GeV}  \\ \hline
            $Z'$ & \multicolumn{2}{c|}{$\mathcal{O}(10)$ MeV} \\ \hline
        \end{tabular}
        \caption{\label{tab:masses} The relevant mass scales for the new particles in the super-weak model. Sterile neutrino $\nu_4$ fulfills two possible dark matter scenarios corresponding to two distinct mass scales. Bounds for $Z'$ mass can given by dark matter scenario \cite{Iwamoto:2021fup}. Masses of the quasi-mass-degenerate neutrinos $\nu_5$ and $\nu_6$ are chosen to combine the sensitivity of them to near-future experiments and the resonant leptogenesis mechanism. Mass of the scalar $s$ is constrained by vacuum stability \cite{Peli:2019vtp}.}
    \end{table}
    
    \subsection{Active-sterile mixing}
    
    We use the Casas-Ibarra parameterization to write the Yukawa matrix $Y_\nu$ in terms of the neutrino mixing matrix elements $U_{ij}^\text{PMNS}$ and neutrino masses $m_1,\dots, m_6$  as
    \be 
    Y_\nu = \frac{\sqrt{2}}{v}U^\text{PMNS,*}\sqrt{M^\text{diag}_L}(-iR^T)\sqrt{M_N}.
    \ee 
    The $3\times 3$ active-sterile mixing matrix can the be written as
    \be 
    U_\text{as} = \begin{pmatrix}
        U_{e4} & U_{e5} & U_{e6} \\ U_{\mu 4} & U_{\mu 5} & U_{\mu 6} \\ U_{\tau 4} & U_{\tau 5} & U_{\tau 6} 
    \end{pmatrix} = m_D^*m_R^{-1\dagger} = U^\text{PMNS}\sqrt{M_L^\text{diag}}iR^T\sqrt{M_R^{-1}}
    \ee 
    
\begin{figure}[htb]
        \centerline{%
            \includegraphics[width=12.5cm]{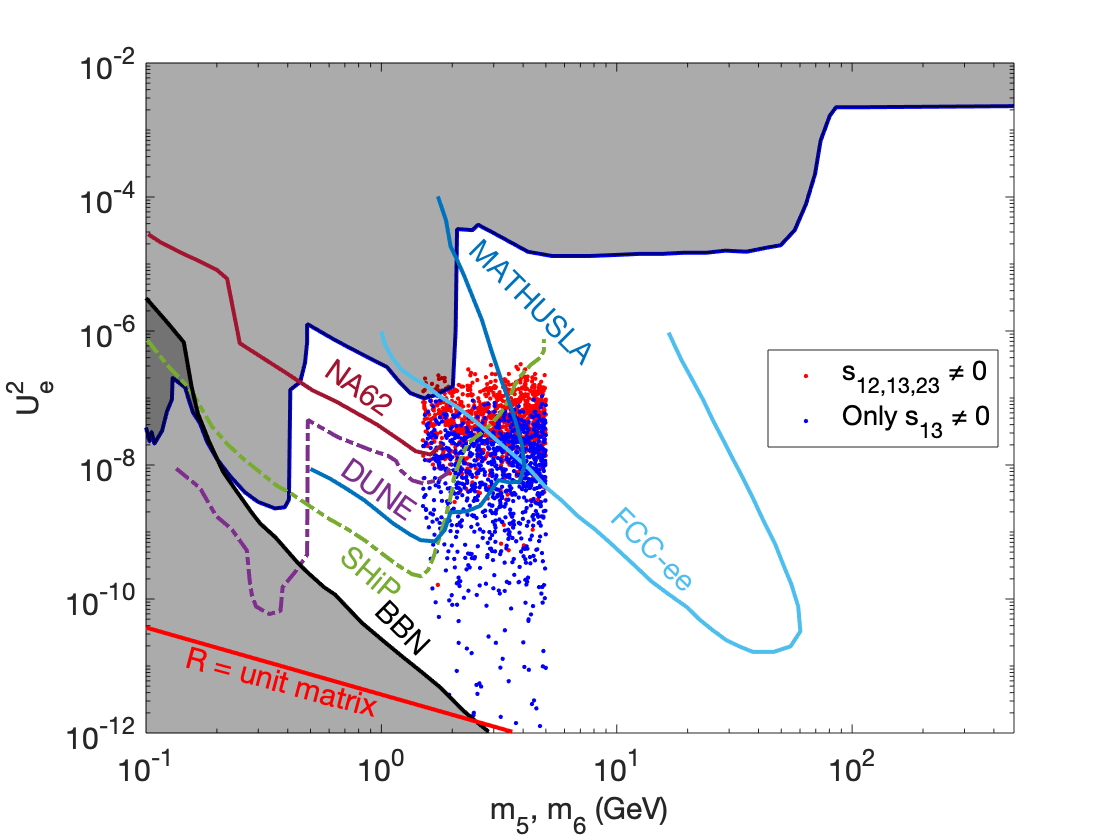}
}
        \caption{Scatter plot of weight of electron flavour of sterile neutrinos in the case where $R$ matrix is free (red dots) and where only $s_{13}$ is nonzero (blue dots). Grayed out area is excluded by current experiments and the coloured lines correspond to expected sensitivities of future experiment. The BBN line corresponds to requirement that the sterile neutrinos $\nu_5$ and $\nu_6$ have lifetime less than 1 s. Default case $R = \textbf{1}_3$ corresponds to the red line. \label{Fig:ScanDemo}}
    \end{figure}

    The matrix $R$ is an arbitrary complex orthogonal matrix. Here we assume that $R$ is real, hence it can be parametrized as
    \be 
    R = \begin{pmatrix}
        c_{12} & -s_{12} & 0 \\ s_{12} & c_{12} & 0 \\ 0 & 0 & 1
    \end{pmatrix}\times \begin{pmatrix}
        c_{13} & 0 & s_{13} & \\ 0 & 1 & 0 \\ -s_{13} & 0 & c_{13}
    \end{pmatrix}\times \begin{pmatrix}
        1 & 0 & 0 \\ 0 & c_{23} & -s_{23} \\ 0 & s_{23} & c_{23}
    \end{pmatrix},
    \ee 
    where $c_{ij} = \sqrt{1-s_{ij}^2}$ and $s_{ij} \in [0,1]$. 
    The usual choice is $R=\textbf{1}_3$, corresponding to the expected mixing from seesaw mechanism. However, this mixing may be enhanced when the matrix $R$ is allowed to vary. This enables us to generated several orders of magnitude larger mixing. To illustrate the expected enhancement by our parametrization, we performed a random scan over $s_{12},s_{13},s_{23} \in [0,1]$, $m_4 \in [10,50]$ keV and $m_5 \in [1.5,5]$ GeV and calculated the weight of electron flavour in sterile states, $U_e^2 \equiv |U_{e4}|^2 + |U_{e5}|^2 + |U_{e6}|^2$. We repeated the scan also in the case where $s_{12} = s_{23} = 0$. The result is in Fig. \ref{Fig:ScanDemo}.
    
    \subsection{Nonstandard interactions}
    
    Consider the tree-level active neutrino - charged fermion scattering process 
    $$
    \nu_\ell + f \rightarrow Z^{\prime*} \rightarrow \nu_\ell + f,\quad f = e,u,d.
    $$
    These processes are manifestations of the super-weak force. For the purpose of analyzing neutrino scattering, the virtual $Z'$ can be integrated out as long as its mass is at least about 10\:MeV, which is the case in our model. The resulting operators have the well-known form of neutral current nonstandard neutrino interactions (NSI). They are a set of effective nonrenormalizable dimension-6 operators of (V$-$A)(V$\pm$A) form,
    \begin{align}
        \lt_\text{NSI} &= -2\sqrt{2}G_F\eps_{\ell\ell'}^{f,C}(\overline{\nu_\ell} \gamma^\mu P_L\nu_{\ell})(\overline{f}\gamma_\mu P_Cf).
    \end{align} 
    Here $G_F$ is the Fermi coupling, $\ell,\ell' = e,\mu,\tau$, $C=L,R$, and $P_C$ are the chiral projection operators. The NSI parameters $\eps^{f,C}_{\ell\ell'}$ are dimensionless and in general complex. In the super-weak model they are real:
    \bal
    \eps^{f,C}_{\ell\ell} &= \frac{v^2}{2M_{Z'}^2}\sul{eC^L_{Z'\nu\nu}}\sul{eC^X_{Z'ff}},
    \eal 
    with chiral couplings given in Eq.~\ref{eq:eC}. We define $\eps^f_{\ell\ell} = \eps^{f,L}_{\ell\ell} + \eps^{f,R}_{\ell\ell}$. The NSI in the super-weak model are flavour-universal: $\eps^f_{ee} = \eps^f_{\mu\mu} = \eps^f_{\tau\tau} \equiv \eps^f$. Summing over the contributions of NSI on neutrino propagation in matter and assuming neutrality of matter, the electron and proton contribution to NSI vanishes due to suitable U(1)$_z$ charge assignments. The resulting NSI is simply \cite{Karkkainen:2021syu}
    \be \begin{split}
        \eps^m = -\frac{v^2}{8M_{Z'}^2}\frac{N_n}{N_e}&\sul{g_y'\cos \theta_Z - \frac{g_L}{\cos \theta_W}\sin \theta_Z}\\
        &\times\sul{(g_y'-g_z')\cos \theta_Z - \frac{g_L}{\cos\theta_W}\sin \theta_Z}.
    \end{split}
    \ee 
    The NSI effect is further suppressed in neutrino oscillation experiments due to active-sterile mixing, but unsuppressed in neutrino scattering expriments. The region defined by $|\eps^m| < \mathcal{O}(0.1)$ is consistent with experimental limits. This region corresponds to $|\theta_Z| < \mathcal{O}(10^{-3})$ and excludes low values of VEV of singlet scalar $\chi$ \cite{Karkkainen:2021syu}.

    \section{Conclusions}
    
    We have demonstrated the experimental feasability of detecting the effects of the neutrino sector in the super-weak model. Neutrino masses are generated -- after spontaneous symmetry breaking of an extra scalar field $s$ -- by type-I seesaw mechanism. We have checked that the one-loop correction to active neutrino masses induced by gauge, scalar and Goldstone bosons are tiny. The active-sterile mixing may be enhanced by several orders of magnitude compared to usual choice of $R=\textbf{1}_3$, allowing the near-future experiments to access a significant region in the parameter space of the model. The neutral current NSI effects may be large, which allows their detection via neutrino scattering experiments. Thus, the super-weak model may be probed independently from multiple different sectors.
    
    \subsection*{Acknowledgments}
    
    We are grateful to Sho Iwamoto for fruitful discussions and to Josu 
    Hernández-García for careful reading of the manuscript. This work was 
    supported by the National Research, Development and Innovation 
    Office -- NKFIH K 125105.
    
    \bibliographystyle{unsrt}
    \bibliography{template.bib}

\begin{thebibliography}{1}

\bibitem{Trocsanyi:2018bkm}
Zoltán Trócsányi.
\newblock {Super-weak force and neutrino masses}.
\newblock {\em Symmetry}, 12(1):107, 2020.

\bibitem{Iwamoto:2021wko}
Sho Iwamoto, Timo~J. K\"arkk\"ainen, Zolt\'an P\'eli, and Zolt\'an
  Tr\'ocs\'anyi.
\newblock {One-loop corrections to light neutrino masses in gauged U(1)
  extensions of the standard model}.
\newblock {\em Phys. Rev. D}, 104(5):055042, 2021.

\bibitem{Iwamoto:2021fup}
Sho Iwamoto, K\'aroly Seller, and Zolt\'an Tr\'ocs\'anyi.
\newblock {Sterile neutrino dark matter in a U(1) extension of the standard
  model}.
\newblock arXiv eprint hep-ph/2104.11248.

\bibitem{Peli:2019vtp}
Zolt\'an P\'eli, Istv\'an N\'andori, and Zolt\'an Tr\'ocs\'anyi.
\newblock {Particle physics model of curvaton inflation in a stable universe}.
\newblock {\em Phys. Rev. D}, 101(6):063533, 2020.

\bibitem{Karkkainen:2021syu}
Timo~J. K\"arkk\"ainen and Zolt\'an Tr\'ocs\'anyi.
\newblock {Experimental constraints on the neutrino and gauge parameters of the
  super-weak U(1) extension of the standard model}.
\newblock arXiv eprint hep-ph/2105.013350.

\end{thebibliography}
    
\end{document}